\documentclass[a4paper,fleqn,natbib]{cas-dc}
\usepackage{graphicx}
\usepackage{caption} % Make sure this package is included
\def\tsc#1{\csdef{#1}{\textsc{\lowercase{#1}}\xspace}}
\tsc{WGM}
\tsc{QE}
\tsc{EP}
\tsc{PMS}
\tsc{BEC}
\tsc{DE}

\begin{document}
\let\WriteBookmarks\relax
\def\floatpagepagefraction{1}
\def\textpagefraction{.001}

% Short title
\shorttitle{Deep Learning for Battery State Prediction}

% Main title of the paper
\title[mode = title]{A Deep Learning Model for Battery State Prediction towards Intelligent Energy Management}                      

% Title footnote 1.
\tnotetext[1]{This document is the results of the research collaboration of iPRISM Research Group and Sunlight Group.}

% Address/affiliation
\affiliation[1]{organization={Department of Informatics and Telecommunications, University of Thessaly},    addressline={3rd km Old National Road},    city={Lamia},    postcode={35100 GR},     country={Greece}}
    
\affiliation[2]{organization={Sunlight Group},   addressline={Thivaidos 22, Kifisia},    city={Athens},   postcode={14564},   country={Athens}}

% First author
\author[1]{Athanasios Koukosias}[type=editor,orcid=0009-0006-9220-8942,prefix=]
% Footnote of the first author
\fnmark[1]
% Email id of the first author
\ead{akoukosias@uth.gr}
%  Credit authorship

% Second author
\author[1]{Vasileios Tzanidakis}[type=editor,orcid=0009-0009-3092-4531,prefix=]
% Footnote of the first author
\fnmark[1]
% Email id of the first author
\ead{vtzanidakis@uth.gr}
%  Credit authorship

% Third author
\author[2]{Sotiris Athanasiou}%[prefix=Dr]
\fnmark[2]
\ead{s.athanasiou@sunlight.gr}

% Fourth author
\author[1]{Kostas Kolomvatsos}[orcid=0000-0002-9442-3340]%[prefix=Dr]
\fnmark[3]
\ead{kostasks@uth.gr}
\ead[URL]{http://kostasks.users.uth.gr}

% Here goes the abstract
\begin{abstract}
Accurate forecasting of battery health indicators, including remaining capacity and lifetime, is of paramount importance for ensuring the reliability, safety, and operational efficiency of applications such as electric vehicles and large scale energy storage infrastructures. The result of the forecasting can be adopted to build an advanced monitoring mechanism for continuous checking batteries' health status to assist in the efficient real-time management of numerous applications. This research investigates the development and implementation of a Deep Learning (DL) model for the prediction of the future state and performance of industrial electrochemical energy storage systems. To address this challenge, we propose a dedicated computational framework that integrates advanced neural network architectures with large-scale training datasets, enabling precise modeling of batteries degradation dynamics and operational trends. The proposed approach provides a decision support mechanism for the optimal management of batteries facilitating both predictive maintenance and the efficient allocation of energy resources. Our findings highlight the potential of DL-based predictive modeling to significantly contribute to the advancement of sustainable and intelligent energy management systems.
\end{abstract}

% Use if graphical abstract is present
% \begin{graphicalabstract}
% \includegraphics{figs/grabs.pdf}
% \end{graphicalabstract}

% Research highlights
%\begin{highlights}
%\item Research highlights item 1
%\item Research highlights item 2
%\item Research highlights item 3
%\end{highlights}

% Keywords
% Each keyword is seperated by \sep

\maketitle

\section{Introduction}
Battery technology is a key enabler for the energy transition, which is a central pillar of the European Green Deal \cite{fetting2020european}. The cornerstone of battery technology is the estimation of batteries' State of Charge (SoC) which is important in all battery powered equipment electric vehicles, stationary storage, robotics, and portable electronics. By monitoring SoC, a Battery Management System (BMS) can make high level decisions such as range prediction for Battery Electric Vehicles (BEVs), power limit enforcement, and charge scheduling. An inaccurate estimation can affect available power, increase the risk of over-discharge, or lead to premature derating, performance losses, and inefficient asset utilization. The more dynamic a system’s duty cycle, the greater the need for accurate SoC prediction. Additionally, since there is a strong connection between lithium-ion battery (LIB) SoC and safety \cite{choi2021analysis}, accurate estimation of this state variable is critical.

There are multiple methods to calculate SoC \cite{xiong2017critical}: conventional SoC estimation methods, like Coulomb counting, Open Circuit Voltage (OCV) lookups, and model based observers  (e.g., Kalman filter families on equivalent-circuit models) \cite{how2019state} perform well in constrained settings. These methods can struggle with OCV hysteresis, sensor drift, temperature dependence, and aging induced parameter shifts. Their accuracy can degrade under partial cycling, fast charging, or rapidly varying loads, and maintaining fidelity over the battery’s life requires repeated re-identification and careful calibration. These challenges are amplified in heterogeneous deployments where chemistries, formats, and usage patterns differ substantially, imposing a heavy engineering burden to sustain reliable SoC across domains.

Deep learning \cite{chemali2017long,zhao2020lithium} offers a compelling alternative by learning expressive mappings directly from time series measurements voltage, current, and temperature without rigid assumptions about linearity or stationarity/non-stationarity. Sequence models (e.g., temporal convolutional networks, LSTMs, and Transformers) can integrate multiscale temporal context, disentangle rate- and temperature-dependent dynamics, and adapt to non-idealities such as hysteresis and noise. When combined with strategies like domain adaptation, physics informed regularization, and uncertainty quantification, neural SoC predictors can generalize across cells and duty cycles while providing calibrated confidence traits essential for safety-critical decision making and optimal energy scheduling.
Although there is a vast amount of battery data available online for this type of work \cite{jin2020energy}, our approach leverages real data from the proprietary, large scale Sunlight Group cloud platform, GLocal \cite{sunlight2025platform}. The Cloud system is responsible for monitoring batteries and ensuring optimal operation, sending real-time notifications upon fault detection. The GLocal system is a powerful tool that enables optimal battery performance for customers while minimizing downtime and maintenance costs. Data from Sunlight owned batteries have been anonymized and processed during this work across multiple use cases.

The key contribution of this work lies in the successful fusion of data driven deep learning approaches with fundamental battery electrochemistry principles. By employing autoencoders for intelligent feature compression and BiLSTM networks for temporal pattern recognition, followed by the application of the Coulomb counting methodology for SoC derivation, the framework leverages the complementary strengths of both methodologies. This hybrid approach not only achieves high prediction accuracy but also maintains physical interpretation ability which is a crucial requirement for real world battery management applications. The consistent performance across multiple battery types and configurations depicted by $R^2$ scores reaching up to $94\%$ underscores the model's robustness and generalization capability.
The following list briefly describes the contributions of our paper:
\begin{itemize}
    \item \textbf{A hybrid SoC estimation framework:} A novel integration of autoencoders, BiLSTM networks, and Coulomb counting, combining data-driven learning with electrochemical consistency;
    \item \textbf{An efficient feature learning model:} Autoencoders compress multi-sensor data into compact latent features, improving learning efficiency and reducing redundancy;
    \item \textbf{Temporal pattern extraction:} BiLSTM captures bidirectional temporal dependencies in charge--discharge sequences for accurate SoC prediction;
    \item \textbf{Experimental validation:} The model is trained and tested on real world datasets from the Sunlight GLocal platform, achieving up to 94\% $R^2$ and demonstrating strong robustness across battery types.
\end{itemize}

The remainder of the paper is organized as follows. Section~\ref{sec:relatedwork} reviews related work on battery state prediction using data-driven methods. Section~\ref{sec:preliminary} introduces the battery systems and the data model used for analysis. Section~\ref{sec:forecasting} describes the forecasting methodology, including SVR and LSTM-based architectures, with a focus on the hybrid Autoencoder-BiLSTM framework for SoC prediction. Section~\ref{sec:experiments} presents the experimental evaluation, including datasets, performance metrics (MAE, RMSE, \(R^2\)), and results. Finally, Section~\ref{sec:conclusion} concludes the paper and discusses future research directions.

\section{Related Work}
\label{sec:relatedwork}
Data driven and deep learning approaches have become central to SoC estimation, offering a powerful alternative to equivalent circuit models. Among these, recurrent neural networks particularly LSTM and Gated Recurrent Unit (GRU) architectures have demonstrated strong capability in learning nonlinear dependencies from current, voltage, temperature and other measurements. 
Early work established the feasibility of LSTM-based SoC estimation, achieving sub-percent accuracy under controlled temperature conditions \cite{chemali2017}. Later methods exploited hybrid convolutional recurrent structures to combine spatial and temporal features, improving prediction accuracy across dynamic automotive profiles \cite{song2017}. 
Bidirectional recurrent networks further enhanced estimation robustness by processing input sequences in both forward and backward directions, yielding improved generalization under diverse load conditions \cite{zhang2020}. 
GRU-based estimators have gained attention due to their reduced computational complexity and faster convergence. Adaptive GRU frameworks with online learning rate adjustment have demonstrated superior performance compared to traditional RNN approaches on real LiFePO\textsubscript{4} datasets \cite{javid2020}. Deeper GRU architectures have been also evaluated on commercial lithium-ion cells, achieving MAE values below 2\% \cite{hassan2023}. For embedded applications, lightweight GRU models with engineered features have achieved high accuracy (MAE $\approx$ 0.85\%) while maintaining low computational requirements \cite{uday2022}. Optimized bidirectional GRU (BiGRU) networks have also been applied to joint SOC and State of Energy (SoE) estimation under broad operating conditions \cite{chen2024}. Beyond recurrent models, several hybrid and emerging architectures have been proposed. Pipelined recurrent networks have been used for simultaneous SoC and voltage prediction \cite{capizzi2011}. Transformer-based models, benefiting from global attention mechanisms, have shown improved long-term dependency modeling and smoother SOC trajectories relative to LSTM baselines \cite{shen2022}. More recent work introduced parallel LSTM structures designed to model charge, discharge, and idle states separately, achieving MAE values near 0.75\% on automotive drive cycles \cite{ozer2025}. Overall, existing literature reveals three key trends: 
 \begin{enumerate}
     \item[-] Temporal sequence learning through LSTM/GRU architectures has become the dominant paradigm for data-driven SOC estimation;
     \item[-] Hybridization via convolutional front-ends, bidirectional processing, or Transformer encoders has improved model robustness under varying thermal and load conditions;
     \item[-] Most studies rely primarily on laboratory datasets, limiting their direct applicability to real-world battery systems.
 \end{enumerate}   
To address this final gap, the present work trains and validates an LSTM-based SOC estimator using operational data collected from industrial battery systems. This contributes to bridging the divide between academic advancements and practical deployment in Battery Management Systems (BMSs), supporting reliable and scalable energy management applications.

\begin{comment}
\section[Theorem and ...]{Theorem and theorem like environments}
{cas-dc.cls} provides a few shortcuts to format theorems and
theorem-like environments with ease. In all commands the options that
are used with the \verb+\newtheorem+ command will work exactly in the same
manner. {cas-dc.cls} provides three commands to format theorem or
theorem-like environments: 

\begin{verbatim}
 \newtheorem{theorem}{Theorem}
 \newtheorem{lemma}[theorem]{Lemma}
 \newdefinition{rmk}{Remark}
 \newproof{pf}{Proof}
 \newproof{pot}{Proof of Theorem \ref{thm2}}
\end{verbatim}

The \verb+\newtheorem+ command formats a
theorem in \LaTeX's default style with italicized font, bold font
for theorem heading and theorem number at the right hand side of the
theorem heading.  It also optionally accepts an argument which
will be printed as an extra heading in parentheses. 

\begin{verbatim}
  \begin{theorem} 
   For system (8), consensus can be achieved with 
   $\|T_{\omega z}$ ...
     \begin{eqnarray}\label{10}
     ....
     \end{eqnarray}
  \end{theorem}
\end{verbatim}  

\newtheorem{theorem}{Theorem}

\begin{theorem}
For system (8), consensus can be achieved with 
$\|T_{\omega z}$ ...
\begin{eqnarray}\label{10}
....
\end{eqnarray}
\end{theorem}
\end{comment}

\section{Preliminary Information}
\label{sec:preliminary}
%\subsection{Battery Systems}
This work focuses primarily on traction batteries for industrial mobility applications. Traction batteries are low voltage systems operating at 51.2 V with capacities between 360 and 504 Ah, based on lithium iron phosphate (LFP) cathode chemistry.
The LFP prismatic cells are ideal for industrial mobility applications as a result of their high cycle life and the inherent safety of LFP chemistry. Each cell has a capacity of 72 Ah and a cycle life of more than 2,000 cycles. Both industrial mobility and traction battery systems, employ a state of the art actively balanced BMS fully compliant with stringent functional safety requirements. Figures 1 and 2 illustrate the general architecture of these two battery systems.

\captionsetup[figure]{name=Fig.}  % This changes "Figure" to "Fig."
\begin{figure}
    \centering
    \includegraphics[width=1.0 \linewidth]{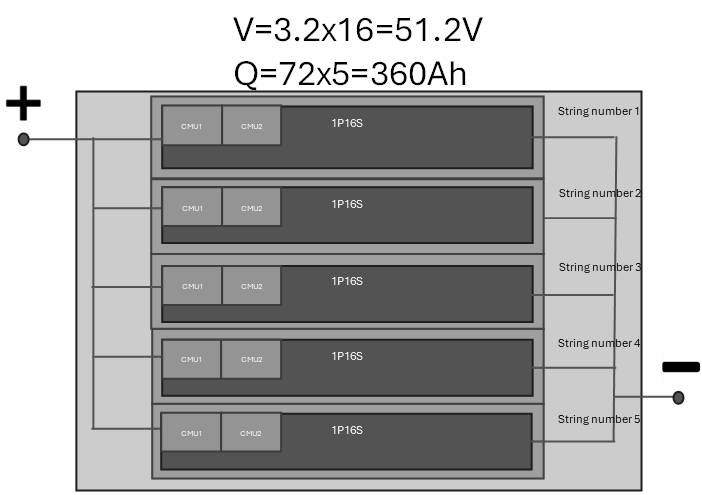}
    \caption{A 51.2V, 360Ah battery system}
    \label{fig:placeholder1}
\end{figure}

\begin{figure}
    \centering
    \includegraphics[width=1.0 \linewidth]{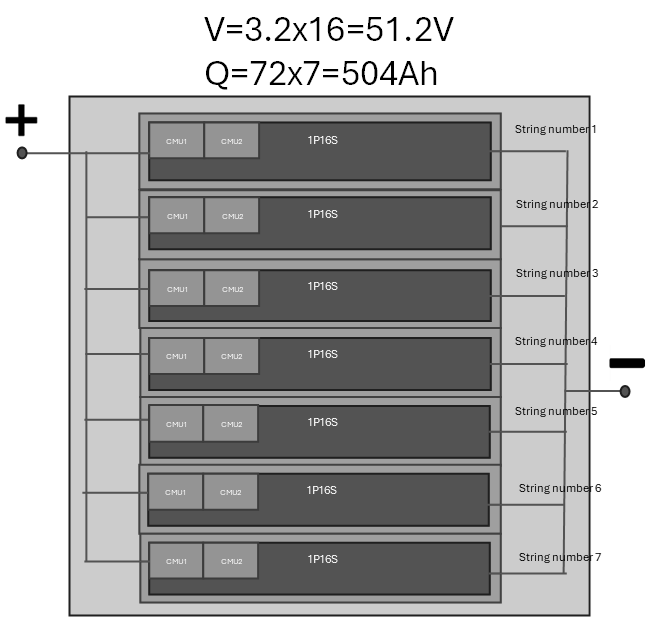}
    \caption{A 51.2V, 504Ah battery system}
    \label{fig:placeholder2}
\end{figure}

%\subsection{Data Model}
The data being analyzed by this study %were jointly prepared through equal collaboration between the industrial partner and the research team. They all 
follow a specific data model which contains operational and diagnostic information collected from battery modules composed of multiple individual but identical cells. Each record represents operational data including a range of electrical, thermal and performance related metrics.
Data are gathered in an industrial environment through the use of a lab monitoring infrastructure built into industrial batteries management system. Measurements capture both cell and module level diagnostics across different operational scenarios which we will later refer to as \textit{User Profiles}. In order to maintain confidentiality and the IPRs of the industrial environment, specific details of the data acquisition methodology and definitions are not disclosed in this work. 
In general, the dataset provides a representative view of the real world performance and usage conditions of batteries formulating a `strong' dataset for training the implemented DL models.

\begin{figure}
    \centering
    \includegraphics[width=1.0\linewidth]{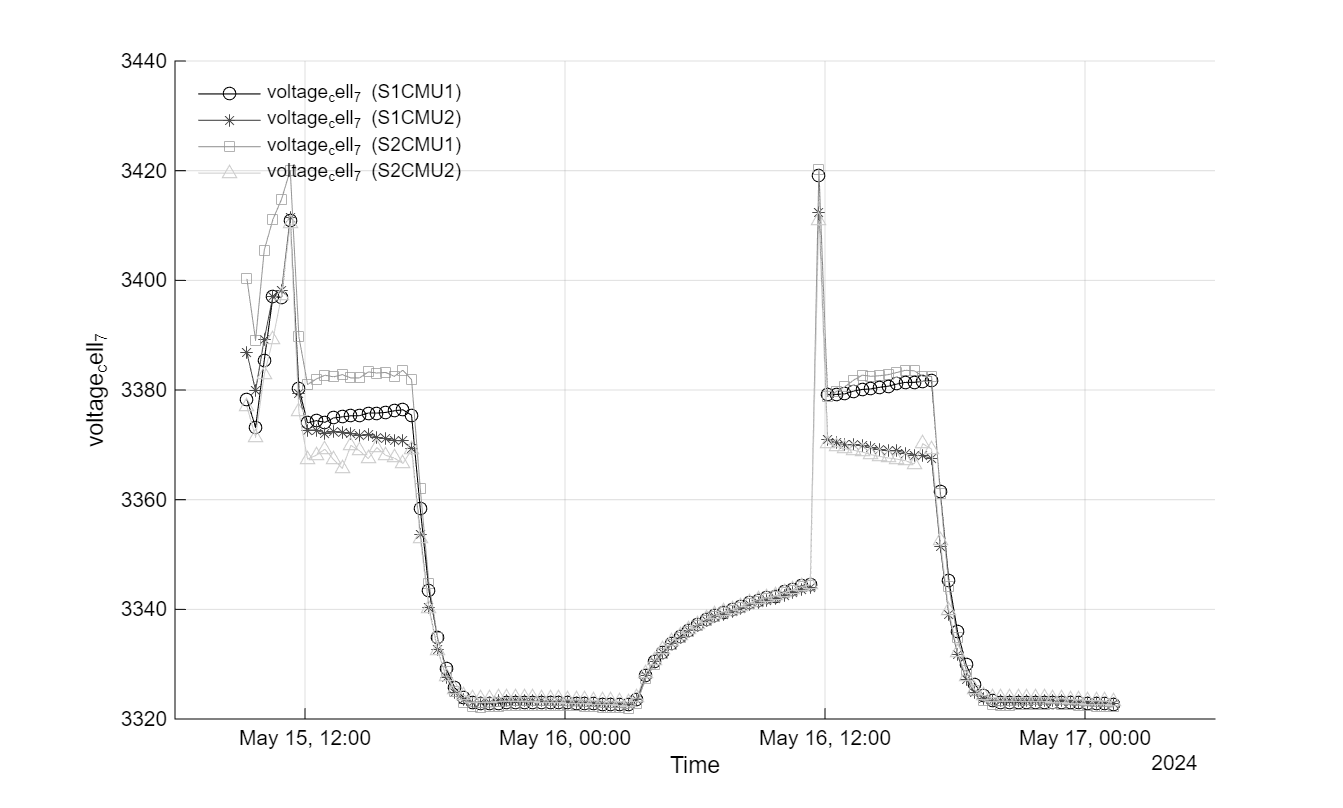}
    \caption{Voltage of the 7th cell per String-CMU}
    \label{fig:placeholder3}
\end{figure}

Prior to data usage, an initial pre-processing stage is applied to prepare the raw sensor outputs for analysis. This step ensured data consistency and anonymization while preserving relationships between the key operational parameters vital to this research.
%After receiving the pre-processed data, the research team performed additional data cleaning and selection techniques. 
In the second phase, recordings containing incomplete values are excluded and only variables relevant to the analysis are retained.
The initial goal is to investigate the behavioral patterns of battery cells and explore potential correlations between performance indicators that could support efficient forecasting. 
Visual analyses are conducted at different levels such as individual cells, modules and battery strings, to further understand how local performance variations affect the overall battery operation. These visualizations set the foundations for the identification of trends, anomalies and imbalances in vital metrics such as voltage, temperature, state of charge and more.
Data includes temporal identifiers, operating mode indicators and a set of thermal variables (see Figures \ref{fig:placeholder3} \& \ref{fig:placeholder4} for a couple of examples). While the precise definitions of these variables remain proprietary, they collectively describe a long term performance state of the modules, forming the basis for subsequent modeling and forecasting stages.

\begin{figure}
    \centering
    \includegraphics[width=1.0\linewidth]{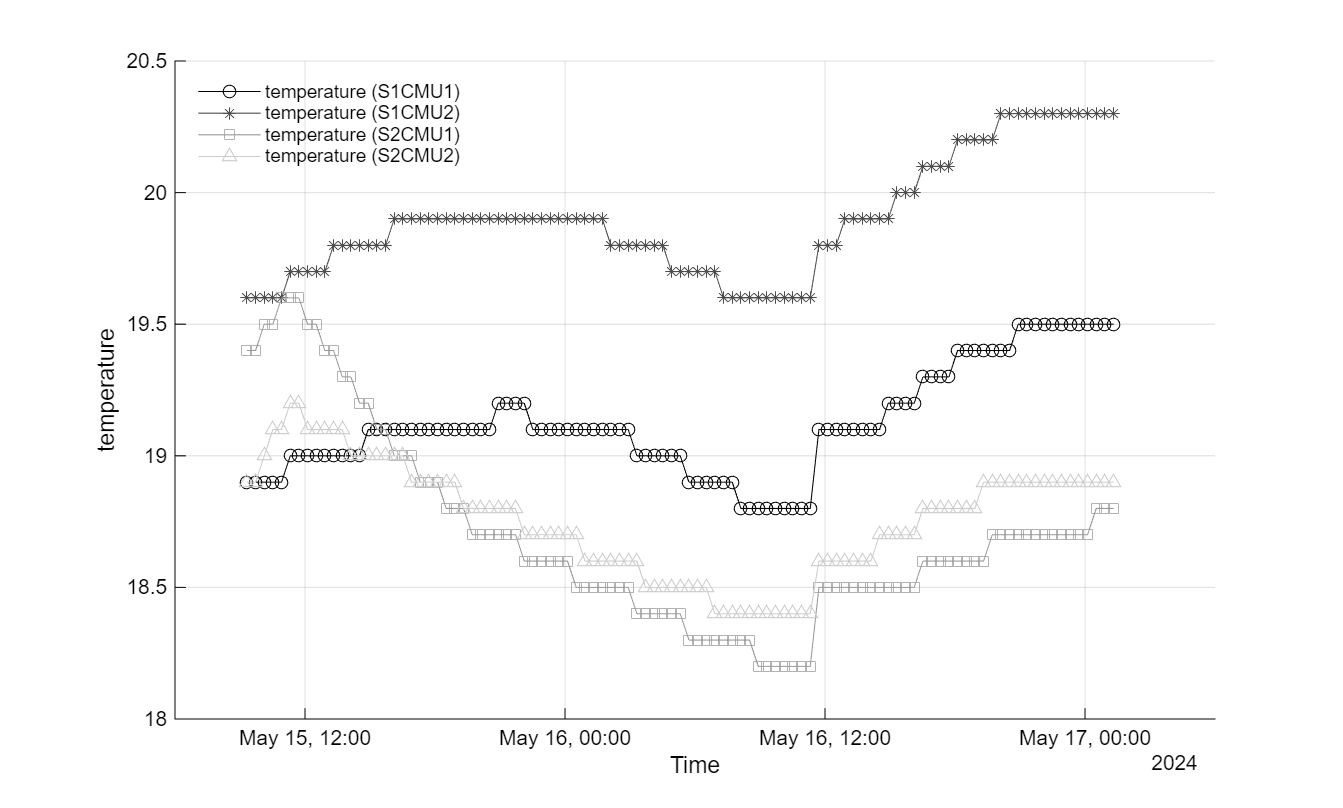}
    \caption{Temperature of a module}
    \label{fig:placeholder4}
\end{figure}

\section{Forecasting the Future State of Batteries}
\label{sec:forecasting}
\subsection{Basic Setup and Model Comparison}
In the initial phase of the work, our objective is to explore data-driven approaches for predicting the SoC of the battery modules using the data provided by an industrial partner. Following a review of the relevant literature and internal reports, which are also referenced in the \textit{Related Work} Section, two machine learning approaches are identified as particularly suitable for this task: \textbf{(a)} Support Vector Regression (SVR) and \textbf{(b)} Long-Short Term Memory (LSTM) network. Both methods have been widely applied in battery management and time series forecasting but differ significantly in their structure and ability to compute and capture time series correlations.

SVR is a regression extension of the Support Vector Machines (SVM) framework, which seeks to identify a function that deviates from the actual targets less than a defined tolerance, while maintaining the model as flat as possible. SVR optimizes the following function:

\begin{equation}
    f(\mathbf{x}) = \mathbf{w}^T \phi(\mathbf{x}) + b
\end{equation} 
where:
\begin{itemize}
    \item[-] \(\mathbf{w}\) is the weight vector controlling the slope of the function,
    \item[-] \(b\) is the bias term (intercept),
    \item[-] \(\phi(\mathbf{x})\) is a mapping of the input features \(\mathbf{x}\) into a higher-dimensional space to enable nonlinear regression,
    \item[-] \(T\) denotes the transpose of \(\mathbf{w}\), which converts the column vector \(\mathbf{w}\) into a row vector so that it can be multiplied with \(\phi(\mathbf{x})\) to produce a scalar output.
\end{itemize}
The goal is to approximate the observed target values \(y_i\) for \(i = 1, \dots, N\), while maintaining a flat function and ignoring errors within a tolerance \(\varepsilon\).
The SVR optimization problem, in its primal form, is defined by
\begin{equation}
\min_{\mathbf{w}, b, \xi_i, \xi_i^*} 
\frac{1}{2} \|\mathbf{w}\|^2 + C \sum_{i=1}^{N} (\xi_i + \xi_i^*),
\end{equation}
subject to
\begin{equation}
    y_i - f(\mathbf{x}_i) \le \varepsilon + \xi_i, \quad
f(\mathbf{x}_i) - y_i \le \varepsilon + \xi_i^*, \quad
\xi_i, \xi_i^* \ge 0,
\end{equation}

where:
\begin{itemize}
\item[-] \(C > 0\) is a regularization parameter controlling the trade-off between the flatness of \(f(\mathbf{x})\) and the tolerance for deviations beyond \(\varepsilon\),
\item[-] \(\xi_i, \xi_i^*\) are slack variables that allow some data points to lie outside the \(\varepsilon\)-insensitive tube,
\item[-] \(\varepsilon\) defines the width of the tube within which errors are not penalized.
\end{itemize}
For nonlinear relationships, a kernel function \(K(\mathbf{x}_i, \mathbf{x}_j)\) is used to compute inner products in the transformed feature space without explicitly calculating \(\phi(\mathbf{x})\). 
In this study, a Radial Basis Function (RBF) kernel is employed, defined as
\begin{equation}
K(\mathbf{x}_i, \mathbf{x}_j) = \exp\big(-\gamma \|\mathbf{x}_i - \mathbf{x}_j\|^2\big),
\end{equation}
where \(\gamma > 0\) controls the width of the Gaussian function. 
This kernel allows the SVR model to capture potential nonlinear relationships among the input features, which include historical measurements of voltage, current, temperature, and operational states such as charging and discharging cycles. Once trained, the model predicts future SoC values given these operational parameters.
Despite the theoretical suitability of SVR for nonlinear regression, two major limitations are observed in practice: 
\begin{itemize}
    \item [--] \textbf{Computational scalability}: The algorithm's performance degrades with large datasets, leading to long training times;
    \item[--] \textbf{Data sensitivity}: The model's accuracy is strongly affected by the quality and completeness of the input data, which introduces additional pre-processing challenges.
\end{itemize}
Given these constraints, the SVR approach is considered less practical for the available data. % and is not developed further.

The second adopted approach is an LSTM network which is a variant of a Recurrent Neural Network (RNN) designed to capture temporal correlations in sequential data. Unlike SVR, which treats each observation independently, LSTM models can leverage patterns across time by maintaining an internal memory. The model receives as input the same sequences of historical battery measurements as the SVR, thus, making a valid comparison of performance. 
The LSTM architecture is composed of a sequence of interconnected components, each one designed to manage the flow of information over time through three key gating mechanisms: 
\textbf{(a)} a \textit{forget gate}, which determines which past information is discarded; \textbf{(b)} an \textit{input gate}, which regulates new information into the component state; and \textbf{(c)} an \textit{output gate}, which controls the information passed forward to generate predictions. 
The LSTM network manages the flow of information through three gating mechanisms that regulate what to remember, update, or discard at each time step. The mathematical formulation of an LSTM cell is expressed as:

\begin{itemize}
\item \begin{equation}f_t = \sigma(W_f [h_{t-1}, x_t] + b_f)
\end{equation}
\item \begin{equation}i_t = \sigma(W_i [h_{t-1}, x_t] + b_i)\end{equation}
\item \begin{equation}\tilde{C}_t = \tanh(W_C [h_{t-1}, x_t] + b_C)\end{equation}
\item \begin{equation}C_t = f_t \odot C_{t-1} + i_t \odot \tilde{C}_t\end{equation}
\item \begin{equation}o_t = \sigma(W_o [h_{t-1}, x_t] + b_o)\end{equation}
\item \begin{equation}h_t = o_t \odot \tanh(C_t)\end{equation}
\end{itemize}

where:
\begin{itemize}
\item[-] $x_t$ represents the input vector at time $t$.
\item[-] $h_t$ the hidden state, and $C_t$ the cell state.
\item[-] $W_f$, $W_i$, $W_C$, and $W_o$ denote the learnable weights.
\item[-] $b_f$, $b_i$, $b_C$, and $b_o$ are the corresponding biases. 
\item[-] The sigmoid function $\sigma(\cdot)$ and the hyperbolic tangent $\tanh(\cdot)$ control how information passes through the network, while $\odot$ indicates element-wise multiplication.
\end{itemize}

\begin{figure}
    \centering
    \includegraphics[width=1.0\linewidth]{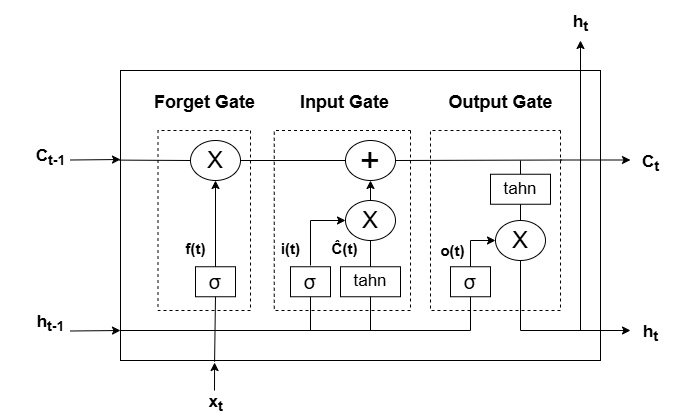}
    \caption{LSTM Architecture}
    \label{fig:lstm}
\end{figure}

As illustrated in Fig.~\ref{fig:lstm}, the \textit{forget gate} $f_t$ determines which past information from $C_{t-1}$ should be retained, the \textit{input gate} $i_t$ decides what new information $\tilde{C}_t$ is added to the cell state, and the \textit{output gate} $o_t$ regulates how much of the internal memory is exposed as the output $h_t$. This gated structure enables the LSTM to effectively capture long-term dependencies, which are essential for accurate State of Charge (SoC) prediction in sequential battery data.

By applying backpropagation through time (BPTT) during training, the network iteratively adjusts its internal weights to capture the correlations between input features and the corresponding SoC values. 

Both methods are initially implemented and evaluated on subsets of the collected dataset. While SVR is significantly a faster method of detecting correlations, it also provides a baseline for non-linear regression. Nevertheless, the LSTM demonstrates superior potential in capturing patterns over time. Additionally, the neural network's flexibility in adapting to complex correlations between variables made it more suitable for long term forecasting tasks. As a result, later development efforts are specifically focused on the LSTM based model, which served as the foundation for the main forecasting framework presented in this study.

\subsection{Advanced LSTM Architectures for Forecasting}
LSTM network, first introduced by Hochreiter and Schmidhuber \cite{graves2012long}, has become a foundational deep learning model for sequential data modeling. Its capability to capture both short and long term temporal dependencies makes it particularly suitable for applications such as time series forecasting, speech recognition, and energy system modeling. 
Within the scope of this study, several LSTM based architectures are explored to identify the configuration that best represents the dynamic behavior of industrial and stationary battery systems.

\subsubsection{Applying a Standard LSTM Model}
The baseline LSTM architecture consists of a single layer with 64 neurons activated by the ReLU function, followed by one or two dense layers with 16-32 neurons using, again, ReLU activation. A final dense layer with a sigmoid activation predicts SoC for each input sequence. Dropout with a probability of 0.2 is applied to reduce overfitting. This model is trained using the Adam optimizer with a learning rate between 0.001 and 0.01. Mean Squared Error (MSE) is used as the loss function, while Mean Absolute Error (MAE) and $R^2$ score serve as additional performance metrics. Training typically involves %10-50 epochs, with 
80\% to 20\% split for training and validation to ensure proper generalization. This architecture achieves $R^2$ scores between $86.7\%$ and $90.3\%$, with predictions further refined using a mean deviation adjustment function to improve alignment with actual SoC measurements.

\subsubsection{An LSTM model combined with an Autoencoder}
\begin{figure}
    \centering
    \includegraphics[width=1.0\linewidth]{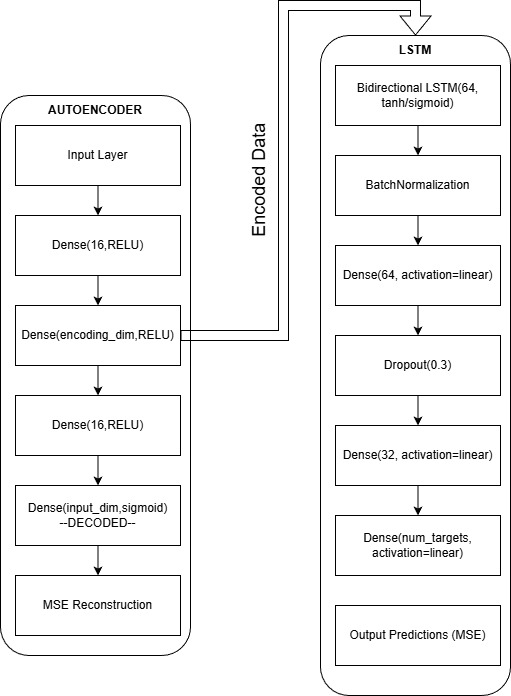}
    \caption{Autoencoder - LSTM Architectures Connection}
    \label{fig:lstm1}
\end{figure}

To improve robustness on new datasets, a pre-trained autoencoder is introduced before data are transferred to the LSTM network (see Figure \ref{fig:lstm1}). 

An autoencoder is a neural network designed to learn compact feature representations by reconstructing input data from a reduced latent space. It consists of two main components: an encoder and a decoder. The encoder maps the input $\mathbf{x} \in \mathbb{R}^m$ to a lower-dimensional latent vector $\mathbf{z} \in \mathbb{R}^n$ $(n < m)$ through:
\begin{equation}
\mathbf{z} = E_{\phi}(\mathbf{x}) = \sigma(\mathbf{W}_e \mathbf{x} + \mathbf{b}_e),
\end{equation}
where:
\begin{itemize} 
\item[-] $\mathbf{W}_e$ and $\mathbf{b}_e$ are the encoder weights and biases
\item[-] $\sigma(\cdot)$ is an activation function. 
\end{itemize}
The decoder reconstructs the input as:
\begin{equation}
\hat{\mathbf{x}} = D_{\theta}(\mathbf{z}) = \sigma(\mathbf{W}_d \mathbf{z} + \mathbf{b}_d),
\end{equation}
with $\mathbf{W}_d$ and $\mathbf{b}_d$ denoting the decoder parameters.
The network is trained to minimize the reconstruction loss, typically defined as:
\begin{equation}
\mathcal{L}_{AE} = \frac{1}{N} \sum_{i=1}^{N} \| \mathbf{x}_i - \hat{\mathbf{x}}_i \|^2,
\end{equation}
which encourages the encoder–decoder pair to capture the most informative features while suppressing noise. The learned latent representation $\mathbf{z}$ is then used as a cleaner, compressed input for the subsequent LSTM model.
The autoencoder compresses input features and detects anomalies allowing the LSTM to process cleaner, more representative data. 
The `hybrid' network architecture remained similar to the baseline LSTM, including dropout at 0.2 and dense layers of 32 and 16 neurons, with the final output using a sigmoid activation for SoC prediction. The integration of the autoencoder enhances the model's generalization capability, particularly when applied to battery usage profiles that differ from the training data, and helps reducing deviations between predicted and actual SoC values. 

\subsubsection{Multi-Output LSTM Architecture}
In scenarios where errors in individual battery cell voltage predictions accumulated and negatively affect SoC estimation, a multi output LSTM is implemented. Instead of predicting the aggregated SoC directly, this architecture predicts the future voltages of each cell individually, which are, then, aggregated to calculate the SoC. Inputs consist of the voltages from all cells, and outputs provide the predicted voltages for each corresponding cell. This approach effectively reduces error propagation and achieves high accuracy with $R^2$ score being between $89\%$ and $94\%$, ensuring reliable SoC estimation across different batches.
Figure \ref{fig:placeholder4} presents an example for voltage prediction based on the proposed architecture.

\begin{figure}
    \centering
    \includegraphics[width=1.0\linewidth]{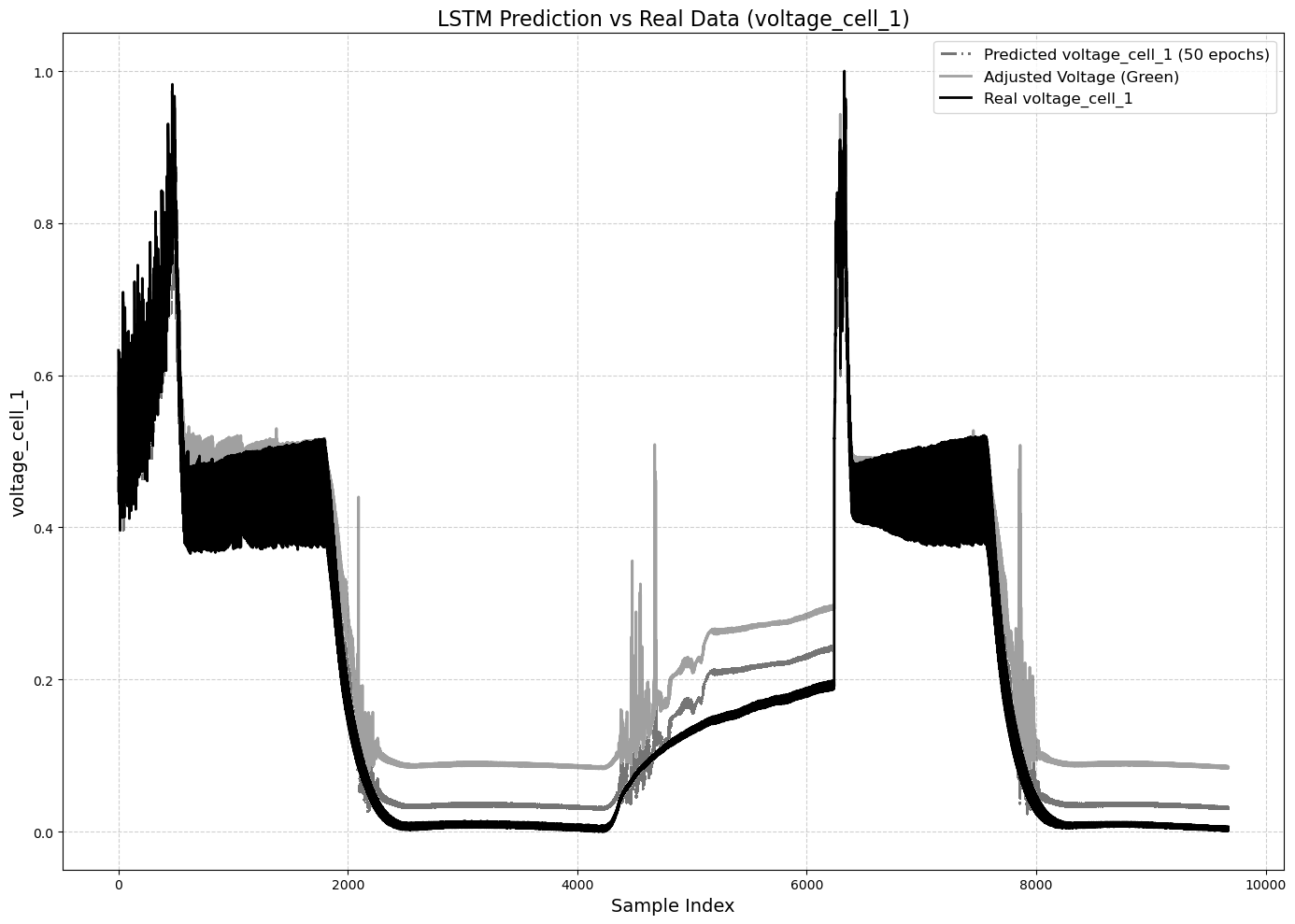}
    \caption{Voltage prediction example of a single battery cell}
    \label{fig:placeholder4}
\end{figure}

\subsubsection{Applying a Bidirectional LSTM (BiLSTM) Architecture}
To capture both past and future temporal dependencies, a Bidirectional LSTM (BiLSTM) architecture is developed. Unlike the standard LSTM which processes sequences in one direction, a BiLSTM consists of two LSTMs: one processing the input forward ($x_1 \to x_T$) and the other backwards ($x_T \to x_1$). For each time step $t$, the forward LSTM computes:
\begin{equation}
\overrightarrow{h}_t = \text{LSTM}_{\text{forward}}(x_t, \overrightarrow{h}_{t-1})
\end{equation}
Here, we have:
\begin{itemize}
\item[-] $x_t$ represents the input vector at the current time step.
\item[-] $\overrightarrow{h}_{t-1}$ is the hidden state from the previous time step in the forward direction.
\item[-] The output $\overrightarrow{h}_t$ is the updated hidden state that encodes information from all previous time steps up to $t$. 
\end{itemize}
Similarly, the backward LSTM computes:
\begin{equation}
\overleftarrow{h}_t = \text{LSTM}_{\text{backward}}(x_t, \overleftarrow{h}_{t+1})
\end{equation}
In this case, we have:
\begin{itemize}
\item[-] $x_t$ is again the input at time step $t$
\item[-] $\overleftarrow{h}_{t+1}$ represents the hidden state from the next time step since the sequence is processed in reverse.
\item[-] The hidden state $\overleftarrow{h}_t$ captures information from all future time steps relative to $t$. 
\end{itemize}
The final hidden state at time $t$ is obtained by concatenating the forward and backward hidden states:
\begin{equation}
h_t^{\text{BiLSTM}} = \overrightarrow{h}_t \, \Vert \, \overleftarrow{h}_t
\end{equation}
,where $\Vert$ denotes concatenation. This combined hidden state incorporates information from both past and future contexts, allowing the BiLSTM to effectively capture complex temporal dependencies in sequential data.

The BiLSTM incorporates a bidirectional layer with 64 neurons per direction, followed by batch normalization and two dense layers of 64 and 32 neurons using ReLU activation. Dropout is defined at 0.3 to prevent overfitting, while the final linear output layer allows regression based predictions. The model is trained using the Adam optimizer with learning rate = 0.001 and MSE loss, a batch size of 16 and 80\% to 20\% training and validation split of the dataset. The BiLSTM significantly improves the prediction of charge current, outperforming previous LSTM variants and effectively capturing complex long term temporal patterns.

\subsection{Final Architecture of the Forecasting Mechanism}
The proposed forecasting system employs a hybrid architecture that integrates an autoencoder for feature compression with a BiLSTM network for temporal modeling \cite{gers2000learning}. This design efficiently processes multivariate battery time series data to forecast future states, with a particular focus on SoC predictions, which is crucial for monitoring the health status of batteries \cite{hochreiter1997long}. 
%The process begins with careful feature selection and normalization. 
Input parameters include SoC, battery voltage, current measurements, and cumulative energy metrics that undergo in a min-max scaling to ensure stable training \cite{hu2012comparative}. 
These normalized features are, then, compressed through a pre-trained autoencoder, creating a compact latent representation that preserves essential information while performing dimensionality reduction to meet computational requirements \cite{how2019state}.

At the core of the forecasting mechanism lies the BiLSTM model \cite{hu2020battery}.
%which, unlike conventional unidirectional models, processes input sequence in both forward and backward directions, enabling richer contextual understanding of battery behavior patterns\cite{torres2021deep}. 
The model's structure incorporates a BiLSTM layer followed by batch normalization, multiple dense layers, and a final multi output layer that simultaneously predicts charge and discharge currents. The training process optimizes the network to minimize prediction error through MSE loss \cite{hannan2017review}. The resulting forecasts are, then, converted to SoC values using the Coulomb counting methodology that tracks net current flow over time, adjusted for battery capacity \cite{cms2024autoencoder}. 
To reflect real world battery management systems behavior, practical rounding logic is applied where SoC values round down during discharge and round up during charging phases \cite{schuster1997bidirectional}.
The complete system supports efficient batch prediction capabilities, enabling processing of extended historical data sequences to generate long term forecasts \cite{al2025machine}. This integrated approach expressiveness of bidirectional networks, resulting in an accurate and practically deployable battery forecasting solution for both operational monitoring and prognostic health management. 

\section{Experimental Evaluation}
\label{sec:experiments}
\subsection{Datasets and Experimental Setup}
The experimental assessment is designed to rigorously validate the model's accuracy, robustness, and generalization capability across diverse battery types and operational scenarios using multiple real-world datasets. These datasets encompass both stationary energy storage systems and traction batteries with varying capacities from 200Ah to 504Ah and different sampling intervals, providing a thorough testing ground under various operational conditions and data characteristics. 
The evaluation methodology employs the chronological splitting of datasets with an 80\% to 20\% ratio for training and validation, respectively, ensuring temporal consistency in testing the model's forecasting capability on unseen data. The model's performance is quantitatively assessed using multiple metrics including Mean Absolute Error (MAE), Root Mean Squared Error (RMSE), and the Coefficient of Determination ($R^2$ score), providing comprehensive insights into both the magnitude and distribution of prediction errors across different battery parameters.
The following equations hold true:
\begin{equation}
MAE = \frac{1}{n} \sum_{i=1}^{n} \big| y_i - \hat{y}_i \big|
\end{equation}

\begin{equation} 
RMSE = \sqrt{\frac{1}{n} \sum_{i=1}^{n} \left( y_i - \hat{y}_i \right)^2}
\end{equation}

\begin{equation}
R^2 = 1 - \frac{\sum_{i=1}^{n} (y_i - \hat{y}_i)^2}{\sum_{i=1}^{n} (y_i - \bar{y})^2}
\end{equation}

where 
\begin{itemize}    
\item[-] $y_i$ is the true value
\item[-] $\hat{y}_i$ is the predicted value
\item[-] $\bar{y}$ is the mean of true values
\item[-] $n$ is the total number of samples. 
MAE measures the average magnitude of prediction errors, RMSE emphasizes larger errors due to squaring, and $R^2$ indicates the proportion of variance explained by the model.
\end{itemize}

\subsection{Performance Assessment}
The results demonstrate consistently high forecasting accuracy when the model is trained and validated on data from the same battery type. In experiments involving traction battery systems, the model achieved $R^2$ scores between $89\%$ and $94\%$ for predicting critical parameters such as ah charged and ah discharged, with the subsequent State of Charge calculations closely tracking measured values. This strong performance was visually collaborated by the close alignment between predicted and actual values in temporal plots, confirming the effectiveness of the integrated Coulomb counting methodology combined with deep learning forecasts. 

\begin{figure*}[htb!]
    \centering
    % Top-most plot
    \includegraphics[width=0.7\textwidth, height=5cm, keepaspectratio]{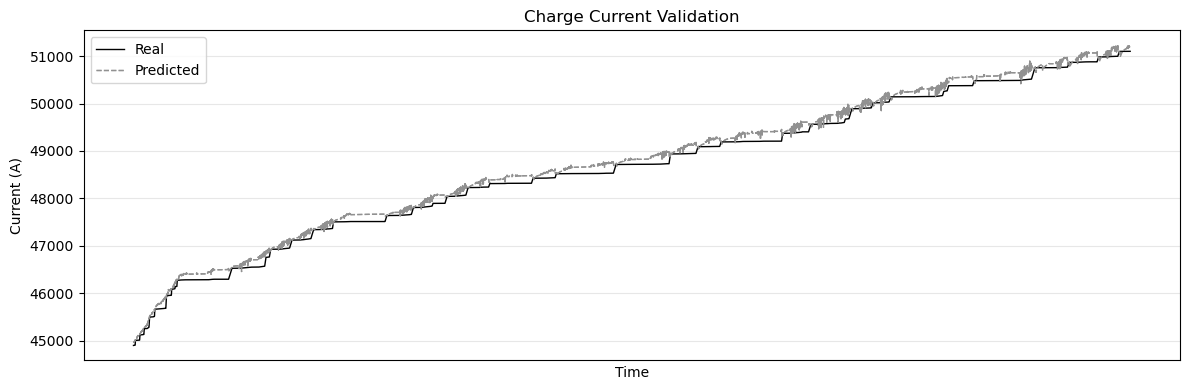}
    \caption{AH. Charge Prediction}
    \label{fig: ah_charged_pred}

    \vspace{0.2cm} % Add vertical space between plots

    % Second plot
    \includegraphics[width=0.7\textwidth, height=5cm, keepaspectratio]{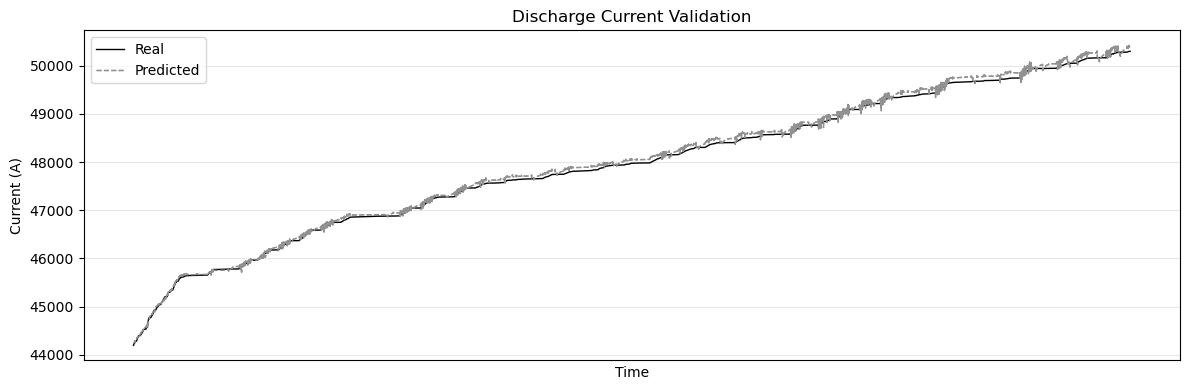}
    \caption{AH. Discharge Prediction}
    \label{fig: ah_discharge_pred}

    \vspace{0.2cm} % Add vertical space between plots

    % Third plot
    \includegraphics[width=1.0\textwidth, height=6cm, keepaspectratio]{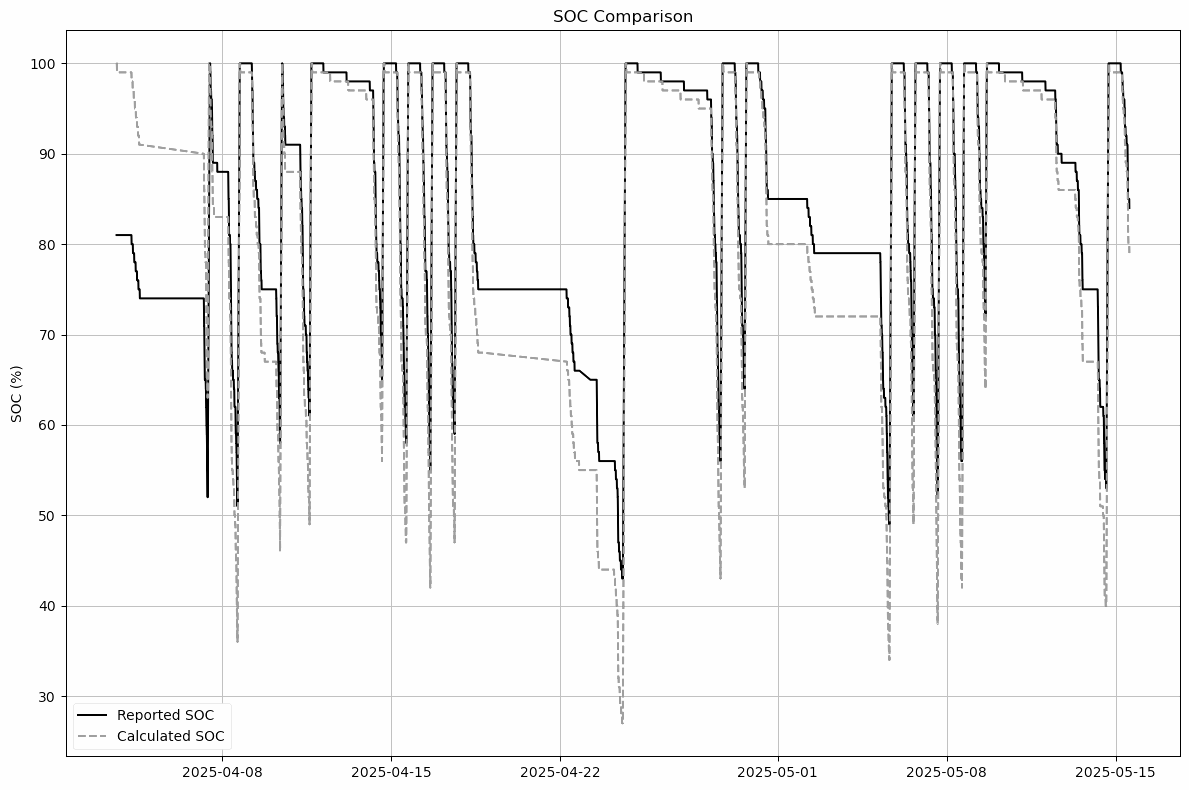}
    \caption{SoC Prediction}
    \label{fig: soc_pred}

    \vspace{0.2cm} % Add vertical space between plots

    % Bottom-most plot
    \includegraphics[width=1.0\textwidth, height=5cm, keepaspectratio]{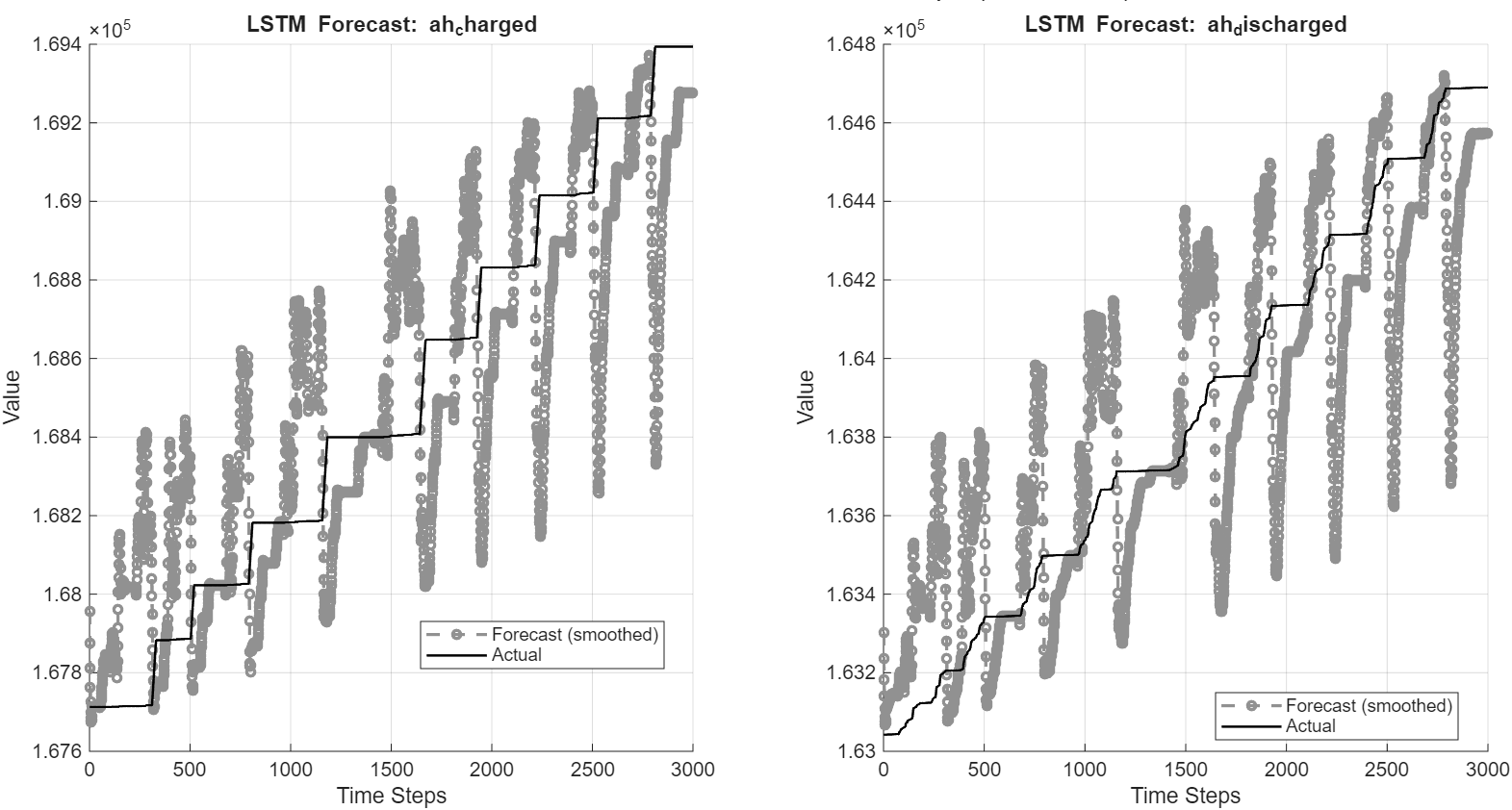}
    \caption{AH. Charge/Discharge Forecasting}
    \label{fig: forecast}

\end{figure*}

The analysis begin with the implementation of the latest version of our LSTM architecture, designed to predict both the ampere-hour (AH) charge and ampere-hour discharge of the battery system. The outcomes of this stage are presented in Figs.~\ref{fig: ah_charged_pred}, ~\ref{fig: ah_discharge_pred}, where the comparison between the predicted (red dashed line) and the measured (blue line) values demonstrates the model’s strong ability to reproduce the real charge–discharge behavior. 
The close agreement between the two curves indicates that the improved LSTM architecture effectively captures the nonlinear temporal dependencies in the data, providing accurate estimations of both charging and discharging trends across the evaluated range.

Using the LSTM-predicted AH charged and discharged values, the Coulomb Counting method is subsequently applied to estimate SoC. The corresponding results are provided by Fig.~\ref{fig: soc_pred}, where the calculated SoC (red dashed line), derived from the LSTM outputs, is compared with the reported SoC (blue line) obtained from on-field measurements. 
The two curves exhibit a high degree of consistency, with deviations appearing mostly during abrupt transitions. This close correspondence validates the reliability of LSTM predictions and confirms that the proposed model provides sufficiently accurate inputs for SoC estimation through Coulomb Counting.

In parallel, the same LSTM derives AH charged and discharged values being employed as inputs to a forecasting model developed to predict future AH charge and discharge behavior. The results of this forecasting stage are illustrated in Fig.~\ref{fig: forecast}, which compares the predicted and measured AH charged and discharged batches. The heatmap represents the deviation between predictions and real data, where greener tones correspond to smaller errors and hence higher accuracy. 
\begin{table*}[htbp]
\centering
\caption{Final architecture performance metrics for various battery configurations \& Forecasting}
\label{tab:battery_metrics}
\renewcommand{\arraystretch}{1.2} % row height
\setlength{\tabcolsep}{10pt} % column spacing
\begin{tabular}{l|c|cc|cc}
\toprule
\textbf{Battery Types} & \textbf{RMSE} & \textbf{MAE (ah\_charged)} & \textbf{$R^2$ (ah\_charged)} & \textbf{MAE (ah\_discharged)} & \textbf{$R^2$ (ah\_discharged)} \\
\midrule
B & 0.083443 & 3008.26 & 0.9270 & 2930.27 & 0.9268 \\
B & 0.044280 & 32.85 & 0.9699 & 27.85 & 0.9717 \\
B $\rightarrow$ EA & 0.080573 & 3075.43 & 0.9237 & 3006.28 & 0.9229 \\
B $\rightarrow$ A & 0.057425 & 1929.55 & 0.9698 & 1863.17 & 0.9702 \\
B $\rightarrow$ B & 0.061759 & 1910.40 & 0.8121 & 1844.63 & 0.8111 \\
B $\rightarrow$ B & 0.082309 & 2612.57 & 0.9221 & 2552.15 & 0.9214 \\
A $\rightarrow$ B & 0.035188 & 40.47 & 0.9836 & 39.46 & 0.9814 \\
A $\rightarrow$ A & 0.099924 & 31.75 & 0.9904 & 29.62 & 0.9907 \\
B $\rightarrow$ A & 0.012225 & 3.83 & 0.9937 & 8.46 & 0.9736 \\
\midrule
Forecast & 272.9508 & 243.9867 & 0.6479 &  207.9981 & 0.7014\\
\bottomrule
\end{tabular}

\vspace{0.5em}
\begin{center}
\textit{B = traction battery B, A = traction battery A, EA = ESS traction battery A}
\end{center}
\end{table*}

The forecasting model achieves accuracy of 64.7\% for AH charged and 70.1\% for AH discharged, demonstrating its ability to anticipate future charging and discharging patterns with consistent performance.

The experimental evaluation collectively demonstrates that the hybrid Autoencoder-BiLSTM architecture provides high accuracy, robustness across battery types, and temporal stability for long term forecasting. The consistent performance across all test scenarios, coupled with efficient batch prediction capabilities, validates the proposed framework as a reliable and practical solution for advanced battery state forecasting in real world energy storage and management applications.

\section{Conclusions}
\label{sec:conclusion}
This research developes and validates a sophisticated hybrid framework for battery state forecasting that integrates autoencoder based feature compression with Bidirectional Long Short Term Memory networks. The comprehensive experimental evaluation demonstrates that the proposed architecture effectively addresses the complex challenge of predicting critical battery parameters, particularly the State of Charge, across diverse battery types and operational conditions. The system's ability to process multivariate time series data while capturing both forward and backward temporal dependencies has proven essential for achieving high forecasting accuracy. 
The experimental results clearly demonstrate the practical viability of the proposed solution for industrial applications. %The model's successful performance in cross type validation scenarios, where it maintained accuracy when applied to battery types not encountered during training, highlights its adaptability to diverse operational environments. 
Furthermore, the temporal stability shown in long term forecasting experiments confirms the framework's suitability for continuous battery health monitoring systems. The batch prediction capability ensures computational efficiency when processing large historical datasets, making the solution scalable for deployment in real world energy management platforms.
Looking forward, this research opens several promising directions for future work. The integration of additional battery health indicators and the extension to probabilistic forecasting could further enhance the framework's utility for predictive maintenance applications. The demonstrated success in cross battery generalization suggests potential for transfer learning approaches that could accelerate deployment in new battery systems with minimal retraining. As the demand for reliable battery management continues to grow across transportation and grid storage applications, the proposed framework provides a solid foundation for developing more intelligent, adaptive, and trustworthy battery analytics systems that can operate effectively in diverse real world conditions.

%\printcredits

\end{document}